\documentclass[submission,copyright,creativecommons]{eptcs}
\providecommand{\event}{BEAT 2014} 

\title{Recursive Session Types Revisited}
\author{Ornela Dardha\footnote
{The author is supported by the UK
EPSRC project {\em From Data Types to Session Types: A Basis for Concurrency and Distribution}
(ABCD) (EP/K034413/1).}
\institute{School of Computing Science, University of Glasgow, UK}
\email{Ornela.Dardha@glasgow.ac.uk}
}
\def\titlerunning{Recursive Session Types Revisited}
\def\authorrunning{O. Dardha}

\usepackage{color}

\usepackage[utf8]{inputenc}
\usepackage[english]{babel}
\usepackage{amsthm,amsmath,amsfonts,amssymb}
\usepackage{txfonts}
\usepackage{proof}
\usepackage{graphicx}
\usepackage{multicol}
\usepackage{listings}
\usepackage{url}
\usepackage{cmll}
\usepackage{algpseudocode}
\usepackage{algorithmicx}
\usepackage{enumerate}
\usepackage{textcomp}
\usepackage{latexsym}
\usepackage{epsfig} 
\usepackage{cite} 
\usepackage{xcolor}
\usepackage{bbm}

\usepackage{packs/mathpartir}
\usepackage{packs/proof-dashed}

\usepackage{makeidx}

\newif{\ifSHORT}
\newif{\iflong}
\newif{\ifLongVersion}
\newif{\ifWithRecords}
\newif{\ifWithProofs}

\newtheorem{theorem}{Theorem}[section]
\newtheorem{lemma}[theorem]{Lemma}
\newtheorem{proposition}[theorem]{Proposition}

\newtheorem{definition}[theorem]{Definition}

\newcommand{\NI}{\noindent}
\newcommand{\smallpar}[1]{\smallskip \NI {\bf #1}}
\newcommand{\m}[1]{\mathsf{#1}}

\newcommand{\wt}[1]{\widetilde{#1}}
\newcommand{\tl}[1]{\tilde{#1}}
\newcommand{\defeq}{\stackrel{\texttt{def}}{=}}

\newcommand{\emp}{\emptyset}

\newcommand{\cmplt}[1]{\mathsf{cplt}(#1)}
\newcommand{\picmplt}[1]{\mathsf{picplt}(#1)} 
\newcommand{\substIO}[2]{\lfloor  ^{#1} / _{#2} \rfloor}

\newcommand{\unfoldSym}{\textsc{unf}}
\newcommand{\unfold}[1]{{\unfoldSym}({#1})}

\newcommand{\st}{\mathtt {\small SType}}
\newcommand{\pt}{\mathtt {\small PType}}

\newcommand{\p}{$\pi$-\! }
\newcommand{\nil}{{\boldsymbol 0}}
\newcommand{\selection}[2]{#1\triangleleft {#2}}

\newcommand{\branching}[3]{#1\triangleright\{{#2}_i:#3_i\}_{i\in I}}

\newcommand{\out}[1]{\oc\langle #1\rangle}
\newcommand{\inp}[1]{\wn({#1})}
\newcommand{\res}[1]{({\boldsymbol \nu} #1)}
\newcommand{\pp}{\ \boldsymbol{|}\ }
\newcommand{\picase}[3]
{\mathbf{case} \, {#1} \, \mathbf{of}\, \{ \mathnormal{l}_i \_ {#2} \triangleright {#3} \}_{i\in I}}
\newcommand{\justcase}{$\mathbf{case}$\ }

\newcommand{\vv}[2]{\mathnormal{l}_{#1}\_ #2}

\newcommand{\variant}[2]{\langle {#1} : #2 \rangle_{i\in I}}

\newcommand{\select}[2]{\oplus\{{#1}_i:#2_i\}_{i\in I}}
\newcommand{\branch}[2]{\&\{{#1}_i:#2_i\}_{i\in I}}

\newcommand{\nilT}{{\lsttxt{end}}}
\newcommand{\rec}[1]{\mu {#1}}

\newcommand{\un}{\m{un}}

\newcommand{\ch}{\sharp}
\newcommand{\ltp}[2]{\ell_{\m{#1}}\ [#2]}

\newcommand{\unit}{{\mathbf{1}}}
\newcommand{ \unitT}{\lsttxt{Unit}}

\newcommand{\sinp}[1]{\wn{#1}.}
\newcommand{\sout}[1]{\oc{#1}.}

\newcommand{\ctx}[3]{\mathcal{#1}^{#2}[#3]}

\newcommand{\mto}{\to^*}

\newcommand{\R}{\mathrel{\ \mathcal{R} \ } }

\newcommand{\sbt}{\ {\preceq_\mathsf{s}}\ }
\newcommand{\sbteq}{\ {=_\mathsf{s}}\ }
\newcommand{\pbt}{\ {\preceq_\mathsf{p}}\ }
\newcommand{\pbteq}{\ {=_\mathsf{p}}\ }

\newcommand{\Sdual}{  \bot_\mathsf{s}\ }
\newcommand{\Pdual}{ \bot_\mathsf{p}\ }

\newcommand{\encf}[1]{\llbracket #1 \rrbracket_{\mathnormal{f}}}
\newcommand{\f}[1]{{\mathnormal f}_{#1}}
\newcommand{\encT}[1]{\llbracket #1\rrbracket}
\newcommand{\enc}[2]{\llbracket #1 \rrbracket_{\mathnormal{f}, \{x,y \mapsto {#2}\}}}
\newcommand{\encx}[2]{\llbracket #1 \rrbracket_{\mathnormal{f},\{x\mapsto {#2}\}}}

\newcommand{\lsttxt}[1]{\text{\rm\tt #1}}

\newdimen\proofrulebreadth \proofrulebreadth=.05em
\newdimen\proofdotseparation \proofdotseparation=1.25ex
\newdimen\proofrulebaseline \proofrulebaseline=2ex
\newcount\proofdotnumber \proofdotnumber=3
\let\then\relax
\def\hfi{\hskip0pt plus.0001fil}
\mathchardef\squigto="3A3B
\newif\ifinsideprooftree\insideprooftreefalse
\newif\ifonleftofproofrule\onleftofproofrulefalse
\newif\ifproofdots\proofdotsfalse
\newif\ifdoubleproof\doubleprooffalse
\let\wereinproofbit\relax
\newdimen\shortenproofleft
\newdimen\shortenproofright
\newdimen\proofbelowshift
\newbox\proofabove
\newbox\proofbelow
\newbox\proofrulename
\def\shiftproofbelow{\let\next\relax\afterassignment\setshiftproofbelow\dimen0 }
\def\shiftproofbelowneg{\def\next{\multiply\dimen0 by-1 }%
\afterassignment\setshiftproofbelow\dimen0 }
\def\setshiftproofbelow{\next\proofbelowshift=\dimen0 }
\def\setproofrulebreadth{\proofrulebreadth}

\def\prooftree{
\ifnum  \lastpenalty=1
\then   \unpenalty
\else   \onleftofproofrulefalse
\fi
\ifonleftofproofrule
\else   \ifinsideprooftree
        \then   \hskip.5em plus1fil
        \fi
\fi
\bgroup
\setbox\proofbelow=\hbox{}\setbox\proofrulename=\hbox{}%
\let\justifies\proofover\let\leadsto\proofoverdots\let\Justifies\proofoverdbl
\let\using\proofusing\let\[\prooftree
\ifinsideprooftree\let\]\endprooftree\fi
\proofdotsfalse\doubleprooffalse
\let\thickness\setproofrulebreadth
\let\shiftright\shiftproofbelow \let\shift\shiftproofbelow
\let\shiftleft\shiftproofbelowneg
\let\ifwasinsideprooftree\ifinsideprooftree
\insideprooftreetrue
\setbox\proofabove=\hbox\bgroup$\displaystyle 
\let\wereinproofbit\prooftree
\shortenproofleft=0pt \shortenproofright=0pt \proofbelowshift=0pt
\onleftofproofruletrue\penalty1
}

\def\eproofbit{
\ifx    \wereinproofbit\prooftree
\then   \ifcase \lastpenalty
        \then   \shortenproofright=0pt  
        \or     \unpenalty\hfil         
        \or     \unpenalty\unskip       
        \else   \shortenproofright=0pt  
        \fi
\fi
\global\dimen0=\shortenproofleft
\global\dimen1=\shortenproofright
\global\dimen2=\proofrulebreadth
\global\dimen3=\proofbelowshift
\global\dimen4=\proofdotseparation
\global\count255=\proofdotnumber
$\egroup  
\shortenproofleft=\dimen0
\shortenproofright=\dimen1
\proofrulebreadth=\dimen2
\proofbelowshift=\dimen3
\proofdotseparation=\dimen4
\proofdotnumber=\count255
}

\def\proofover{
\eproofbit 
\setbox\proofbelow=\hbox\bgroup 
\let\wereinproofbit\proofover
$\displaystyle
}%
\def\proofoverdbl{
\eproofbit 
\doubleprooftrue
\setbox\proofbelow=\hbox\bgroup 
\let\wereinproofbit\proofoverdbl
$\displaystyle
}%
\def\proofoverdots{
\eproofbit 
\proofdotstrue
\setbox\proofbelow=\hbox\bgroup 
\let\wereinproofbit\proofoverdots
$\displaystyle
}%
\def\proofusing{
\eproofbit 
\setbox\proofrulename=\hbox\bgroup 
\let\wereinproofbit\proofusing
\kern0.3em$
}

\def\endprooftree{
\eproofbit 
  \dimen5 =0pt
\dimen0=\wd\proofabove \advance\dimen0-\shortenproofleft
\advance\dimen0-\shortenproofright
\dimen1=.5\dimen0 \advance\dimen1-.5\wd\proofbelow
\dimen4=\dimen1
\advance\dimen1\proofbelowshift \advance\dimen4-\proofbelowshift
\ifdim  \dimen1<0pt
\then   \advance\shortenproofleft\dimen1
        \advance\dimen0-\dimen1
        \dimen1=0pt
        \ifdim  \shortenproofleft<0pt
        \then   \setbox\proofabove=\hbox{%
                        \kern-\shortenproofleft\unhbox\proofabove}%
                \shortenproofleft=0pt
        \fi
\fi
\ifdim  \dimen4<0pt
\then   \advance\shortenproofright\dimen4
        \advance\dimen0-\dimen4
        \dimen4=0pt
\fi
\ifdim  \shortenproofright<\wd\proofrulename
\then   \shortenproofright=\wd\proofrulename
\fi
\dimen2=\shortenproofleft \advance\dimen2 by\dimen1
\dimen3=\shortenproofright\advance\dimen3 by\dimen4
\ifproofdots
\then
        \dimen6=\shortenproofleft \advance\dimen6 .5\dimen0
        \setbox1=\vbox to\proofdotseparation{\vss\hbox{$\cdot$}\vss}%
        \setbox0=\hbox{%
                \advance\dimen6-.5\wd1
                \kern\dimen6
                $\vcenter to\proofdotnumber\proofdotseparation
                        {\leaders\box1\vfill}$%
                \unhbox\proofrulename}%
\else   \dimen6=\fontdimen22\the\textfont2 
        \dimen7=\dimen6
        \advance\dimen6by.5\proofrulebreadth
        \advance\dimen7by-.5\proofrulebreadth
        \setbox0=\hbox{%
                \kern\shortenproofleft
                \ifdoubleproof
                \then   \hbox to\dimen0{%
                        $\mathsurround0pt\mathord=\mkern-6mu%
                        \cleaders\hbox{$\mkern-2mu=\mkern-2mu$}\hfill
                        \mkern-6mu\mathord=$}%
                \else   \vrule height\dimen6 depth-\dimen7 width\dimen0
                \fi
                \unhbox\proofrulename}%
        \ht0=\dimen6 \dp0=-\dimen7
\fi
\let\doll\relax
\ifwasinsideprooftree
\then   \let\VBOX\vbox
\else   \ifmmode\else$\let\doll=$\fi
        \let\VBOX\vcenter
\fi
\VBOX   {\baselineskip\proofrulebaseline \lineskip.2ex
        \expandafter\lineskiplimit\ifproofdots0ex\else-0.6ex\fi
        \hbox   spread\dimen5   {\hfi\unhbox\proofabove\hfi}%
        \hbox{\box0}%
        \hbox   {\kern\dimen2 \box\proofbelow}}\doll%
\global\dimen2=\dimen2
\global\dimen3=\dimen3
\egroup 
\ifonleftofproofrule
\then   \shortenproofleft=\dimen2
\fi
\shortenproofright=\dimen3
\onleftofproofrulefalse
\ifinsideprooftree
\then   \hskip.5em plus 1fil \penalty2
\fi
}

\begin{document}
\maketitle

\begin{abstract}
Session types model structured communication-based programming.
In particular, binary session types for the \p calculus
describe communication between exactly two participants in a distributed scenario.
Adding sessions to the \p calculus means augmenting it with
type and term constructs.
In a previous paper, we tried to understand to which extent the session constructs are
more complex and expressive than the standard \p calculus constructs.
Thus, we presented an encoding of binary session \p calculus to the standard typed \p calculus by adopting
{\em linear} and {\em variant} types and the {\em continuation-passing} principle.
In the present paper, we focus on recursive session types and we present an encoding into recursive linear \p types.
This encoding is a conservative extension of the former in that it preserves the
results therein obtained.
Most importantly, it adopts a new treatment of the duality relation,
which in the presence of recursive types has been proven to be quite challenging.
\end{abstract}

\section{Introduction}

Session types are a type formalism used to model structured communication-based programming for distributed systems.
In particular, binary session types for the \p calculus
describe communication between exactly two participants in such scenario~\cite{THK94,HVK98,GH05,V12}.
%
When sessions are added to the standard typed \p calculus,
the syntax of types and terms is augmented with ad-hoc constructs,
added
on top of the already existing ones.
This yields a duplication of type and term constructs, e.g. restriction of session channels and restriction of standard \p channels
or, recursive session types and recursive standard \p types~\cite{GH05}.
Most importantly, this redundancy is also propagated in the theory of session types:
various properties are proven for session types as well as for standard \p types.
In a previous work~\cite{DGS12}, we focused on a subset of binary session types, namely the finite ones,
and posed the following question:
\begin{center}
{\em To which extent session constructs are more complex and more expressive\\
than the standard \p calculus constructs?}
\end{center}
We answered this question by showing
an encoding of
finite binary session types into
finite {\em linear} \p types
and of
finite session processes into finite standard \p processes.
In the present paper, we extend the encoding
to an infinite setting, namely to recursive session types and replicated processes,
and pose the same question.
We encode recursive session types into recursive linear \p types and
replicated session processes into replicated standard \p processes.
We show that
the current encoding
$i)$ is sound and complete with respect to typing derivations,
intuitively meaning: ``a session process is well-typed if and only if its encoding is well-typed'';
and
$ii)$
satisfies the operational correspondence property,
intuitively meaning: ``a session process and its encoding reduce to processes still related by the encoding''.

The interest and benefits of this encoding are mainly in {\em expressivity} and {\em reusability}
for a larger setting than the one adopted in~\cite{DGS12}.
The encoding is an expressivity result for recursive types.
Its faithfulness, proved by $i)$ and $ii)$,
permits reusability of already existing theory for standard typed \p calculus:
e.g., subject reduction or type safety for session \p calculus
can be obtained as corollaries from the encoding and the corresponding properties in the standard typed \p calculus.

The present encoding is not just an extension of the former,
it presents novelties and differences with respect to~\cite{DGS12}, as listed below.
{\em Duality} is a fundamental notion of session types, as it describes
compatible behaviours between communicating parties.
The most used duality is the {\em inductive duality function} $\ \overline \cdot \ $~\cite{HVK98,V12}.
Recent work~\cite{BH13}
has shown the inadequacy of $\ \overline \cdot \ $ in the presence of recursive types,
because it does not commute with unfolding.
As a consequence, using relations like subtyping or type equivalence 
becomes challenging, because these relations
explicitly use unfolding of recursive types.
In the light of such discovery,
the present encoding
adopts the {\em complement function} $\cmplt$ defined in~\cite{BH13}, which is shown to be adequate,
instead of $\overline \cdot$, adopted in the former encoding.
Since  $\cmplt$ and  $\overline \cdot$  coincide for finite session types,
the encoding in~\cite{DGS12} remains sound and
the present encoding is
a conservative extension of the former,
in that it preserves all the properties that the former encoding satisfies.
For completeness, the present one
is extended to standard variables and hence non session \p processes:
in this case the encoding is an homomorphism and no linearity is required.
On top of $\cmplt$,
we present the {\em co-inductive duality relation}, which is shown to contain the complement~\cite{BDGK14},
and is used in the type system for session \p calculus~\cite{GH05,D14Ext}.
This permits us to give a definition of
complement and co-inductive duality for linear \p calculus types,
which is another contribution of the present paper.

\smallpar{Structure of the paper.}
In \S~\ref{sec:model} we present
the syntax of types and terms for both
the session \p calculus and the standard types \p calculus.
In \S~\ref{sec:encoding} we present the encoding of
recursive session types and session processes and we
state the main results for the encoding.
In \S~\ref{sec:enc_example} we give a detailed example of the encoding
of a well-typed replicated process which uses recursive session types.
We conclude in \S~\ref{sec:concl}.
The proofs of the result herein presented can be found in the online version of the paper~\cite{D14Ext}.


\section{The Model}
\label{sec:model}
\subsection{Background on \p calculus with sessions}
\label{chap:sessions_background}

\smallpar{Syntax.}\label{sec:proc_sessions}
\begin{figure}[h!]
  \begin{displaymath}
    \begin{array}{l}
      \begin{array}{rcllllllll}
        P,Q&  ::= 
        & x\out v.P & \mbox{(output)}
        &\quad | & x\inp y.P & \mbox{(input)}     
        &\quad |  & \selection xl_j.P & \mbox{(selection)}\\
        && \branching xlP & \mbox{(branching)}
        &\quad |  & P \pp Q & \mbox{(parallel)}
        & \quad |  & \res {xy}P & \mbox{(session res.)} \\
        && *P & \mbox{(replication)}
	& \quad |  &\res x P & \mbox{({channel res.})}
        & \quad |   & \nil & \mbox{(inaction)}\\[1mm]
        v  & ::= & x & \mbox{(variable)} & 
       \quad | & \unit & \mbox{(unit value)}
      \end{array}
    \end{array}
  \end{displaymath}
  \vspace{-1.5em}
  \caption{$\pi$-calculus with sessions, syntax.}
  \label{fig:sessionpi}
\end{figure}
The syntax of the $\pi$-calculus with session types~\cite{V12,GH05} is given in Figure~\ref{fig:sessionpi}.

\noindent
$P,Q$ range over processes, $x,y$ over variables, $v$
over values and $l$ over labels.
A value is a variable or $\unit$.
A process is 
	an output $x \out v.P$ which sends $v$ on $x$ with continuation $P$;
	an input $x\wn(y).P$ which receives a value on $x$ and proceeds as $P$;
	a selection $\selection x{l_j}.P$ which selects $l_j$ on $x$ and proceeds as $P$;
	a branching $\branching xlP$ which offers a set of labelled processes on $x$, with labels being all different;
	a parallel composition $P\pp Q$ of $P,Q$;
	replicated $*P$ which spawns copies of $P$;
	a session restriction $\res{xy} P$ or
	a standard channel restriction $\res x P$;
	or $\nil$, the terminated process.
Session restriction differs from the standard one:
$\res{xy}$ states that $x$ and $y$, called {\em co-variables}, are the opposite endpoints of a session channel and are bound in $P$.
It models session creation and the connection phase~\cite{THK94,HVK98}.

\smallpar{Session types.}\label{sec:session_types}
\begin{figure}[h!]
  \begin{displaymath}
    \begin{array}{l}
      \begin{array}{rcllllllll}
        S & ::=
        &\oc T.S & \mbox{(send)}
        & \ |  & \wn T.S& \mbox{(receive)}  
        & \ | & \select lS & \mbox{(select)}\\
        &&\branch lS & \mbox{(branch)} 
        &  \ | & X & \mbox{(type var.)}
        & \ |  &\overline X& \mbox{(dual type var.)}\\
        && \nilT & \mbox{(termination)}
        & \ |  &\rec X.S& \mbox{({rec. session type})}\\[1mm]
        T  & ::= 
        & S & \mbox{({session type})}
        &\ | & \ch T & \mbox{(channel type)}
        & \ | & X & \mbox{(type variable)}\\
        & &\rec X.T& \mbox{({recursive type})}
        &\ |& \unitT & \mbox{(unit type)}
      \end{array}
    \end{array}
  \end{displaymath}
    \vspace{-1.5em}
  \caption{$\pi$-calculus with sessions, types.}
  \label{fig:session_types}
\end{figure}
The syntax of types for the \p calculus with sessions~\cite{GH05} is given in Figure~\ref{fig:session_types}.

\noindent
$S$ ranges over session types and
$T$ over types.
A session type can be
	$\oc T.S$ or $\wn T.S$ which respectively,
	sends or receives a value of type $T$ and continuation $S$; 
	select $\select lS$ or branch $\branch lS$ which are sets of labelled
	session types indicating respectively, internal and external choice, with labels being all different;
	a (dualised) type variable $X, \overline X$, or recursive session type $\rec X.S$
	or the  terminated type $\nilT$.
A type can be
	a session type $S$;
	a standard channel type $\ch T$;
	a type variable $X$ or recursive type $\rec X.T$ or
	a unit type $\unitT$.
Recursive (session) types are required to be {\em guarded}, meaning that
in $\rec X.T$, variable $X$ may occur free in $T$ only under at least one of the other type constructs.
To work with recursive types we need the {\em unfolding} function ($\unfoldSym$) which unfolds
a recursive type until the first type constructor different from $\rec X$ is reached (see~\cite{D14Ext}).
Finally, we use $\st$ to denote the set of
	{\em closed} (no free type variables) and
	{\em guarded} session types.
%

\smallpar{On duality for session types.}
Below we give an adaptation of the complement function~\cite{BH13} to  $X, \overline X$.
\begin{definition}[Complement function for session types]
\label{fig:complement}
\label{def:dual}
The complement function is defined as:
\[
\begin{array}{lcl@{\hskip 1em}lcl}
\cmplt{\wn T.S} & = & \oc{T}.{\cmplt{S}}
&
\cmplt{X} & = & \overline X \\

\cmplt{\oc{T}.{S}} & = & \wn{T}.{\cmplt{S}} 
&
\cmplt{\overline X} & = & X \\

\cmplt{\branch lS} & = & \select l{\cmplt S} 
&
\cmplt{\rec X.S} & = & \rec X.{\cmplt{S\substIO{\rec{X}.{S}}{X}}} \\

\cmplt{\select lS} & = &
\branch l{\cmplt S}&
\cmplt{\nilT} & = & \nilT 
\end{array}
\]
\end{definition}
It uses a syntactic substitution
$\substIO{-}{-}$, which acts only on carried types and is formally defined in~\cite{BH13, BDGK14,D14Ext}.
%
Below we give the definition of standard type substitution for (dualised) type variables~\cite{G08}.
\begin{displaymath}
    \begin{array}{l}
      \begin{array}{lclllll}
        X\{S/X\} 	 		=  S&   			&&Y\{S/X\}  =  Y					& \mbox{ If $X\neq Y$}   \\
        \overline X\{S/X\} 	=  \cmplt S&		&& \overline{Y} \{S/X\} = \overline{Y} 	& \mbox{ If $X\neq Y$}
      \end{array}
    \end{array}
\end{displaymath}
However, when describing opposite behaviours between communicating parties,
in this paper we adopt
the {\em co-inductive duality relation} $\Sdual$ by following~\cite{GH05}.
The benefits of this approach are:
$i)$ $\Sdual$ commutes with unfolding~\cite{BH13} and hence it is adequate;
$ii)$ as stated in~\cite{GH05}, since it is a relation it captures dual behaviours that $\overline\cdot, \cmplt{}$ do not capture,
like
$\rec{X}.{\wn{\unitT}.{X}}$ and
$\oc{\unitT}.{\rec{X}.{\oc{\unitT}.{X}}}$. 
$iii)$ as stated in~\cite{BDGK14}, it contains $\cmplt{}$.
Before defining $\Sdual$, we need the notion of type equivalence  and hence subtyping.
For simplicity, we omit subtyping on base types and on standard channel types, which are given in~\cite{GH05,BH13}.

\begin{definition}[Subtyping and type equivalence for session types~\cite{GH05}]
\label{def:subtyping}
%
A relation $\R \subseteq \st \times \st$ is a {\em type simulation} if $(T,S) \in \R$ implies the following:\\
$i)$ if $\unfold{T}=\nilT$ then $\unfold{S}=\nilT$\\
$ii)$ if $\unfold{T}=\sinp{T_m}T'$ then
			$\unfold{S}=\sinp{S_m}S'$ and $T_m \R S_m $ and
			$T' \R S'$\\
$iii)$ if $\unfold{T}=\sout{T_m}T'$ then
			$\unfold{S}=\sout{S_m}S'$ and $S_m \R T_m$ and
			$T' \R S'$\\
$iv)$ if $\unfold{T}=\branch lT$ then
			$\unfold{S}=\& \{l_j:S_j\}_{j\in J}$,  $I  \subseteq  J$, $T_i \R S_i$, $\forall i \in I$\\
$v)$ if $\unfold{T}=\select lT$ then
			$\unfold{S}=\oplus \{l_j:S_j\}_{j\in J}$,  $J  \subseteq  I$,  $T_j \R S_j$, $\forall j\in J$\\
%
%
%
The {\em subtyping relation} $\sbt$ is defined by $T\sbt S$ if and only if there exists a type simulation $\R$
such that $(T,S) \in \R$.
The {\em type equivalence relation} $\sbteq$ is defined by $T\sbteq S$ if and only if
$T\sbt S$ and $S \sbt T$.
\end{definition}
\begin{definition}[Co-inductive duality for session types~\cite{GH05}]
\label{def:coinductiveduality}
\label{def:Sdual}
A relation $\R \in \st \times \st$
is a {\em duality relation} if $(T,S)\in \R$ implies the following conditions:\\
$i)$ 	If $\unfold{T} = \nilT$ then $\unfold{S} = \nilT$\\
$ii)$ 	If $\unfold{T} = \,\sinp{T_m}{T'}$ then 
		$\unfold{S} = \,\sout{S_m}{S'}$ and $T' \R S'$ and $ T_m \sbteq S_m$\\
$iii)$ If $\unfold{T} = \,\sout{T_m}{T'}$ then 
		$\unfold{S} = \,\sinp{S_m}{S'}$ and $T' \R S'$ and $T_m \sbteq S_m$\\
$iv)$ 	If $\unfold{T} =
		\branch lT$
		then 
		$\unfold{S} =\select lS$ and
		$\forall i \in I$, $T_i \R S_i $ \\
$v)$ If $\unfold{T} =
		\select lT$ then
		$\unfold{S} =
		\branch lS$ and
		$\forall i \in I$, $T_i \R S_i $\\
The {\em co-inductive duality relation} $\Sdual$ is defined by $T\Sdual S$iff
$\ \exists\ \R$, a duality relation such that  $(T,S) \in \R$.
\end{definition}
\begin{proposition}\label{prop:cmplt_dual}
Let $T, S\in \st$, $\cmplt{T} = S\ \Longrightarrow\ T \Sdual S$.
\end{proposition}
%
%
\begin{proposition}[Idempotence]\label{prop:idempotence}
Let $T,S,U\in \st$. If $T\Sdual S$ and $S\Sdual U$ then $T \sbteq U$.
\end{proposition}
By Proposition~\ref{prop:cmplt_dual} and Proposition~\ref{prop:idempotence} we have the following.
\begin{proposition}\label{prop:cmplt_idem}
Let $T,S,U\in \st$. If $\cmplt T = S$ and $\cmplt S = U$ then $T \sbteq U$.
\end{proposition}
%

\subsection{Background on standard \p calculus}

\smallpar{Syntax.}\label{sec:pi_process}
\begin{figure}[h!]
  \begin{displaymath}
    \begin{array}{l}
      \begin{array}{rcllllllll}
        P,Q & ::= 
        & x\out {\tl v}.P & \mbox{(output)}
        &\ | &  x\inp {\tl y}.P & \mbox{(input)}     
        &\ | & P \pp Q & \mbox{(parallel)}\\
        & &\res x P & \mbox{({channel res.})}\
        &\ | & *P & \mbox{(repl.)}
        & \ |  & \picase{v}{x_i}{P_i} & \mbox{({case})}\\
        && \nil & \mbox{(inaction)}\\[1mm]
        v  & ::= 
        & x & \mbox{(variable)}
        &\ | & \unit & \mbox{(unit val.)}
        &\ | & \vv {}v & \mbox{(variant val.)}
      \end{array}
    \end{array}
  \end{displaymath}
  \vspace{-1.5em}
  \caption{Standard $\pi$-calculus, syntax.}
  \label{fig:pi}
\end{figure}
The syntax of the polyadic \p calculus~\cite{Sangio01} is given in Figure~\ref{fig:pi}.

\noindent
$P,Q$ range over processes, $x,y$ over variables, $l$
over labels and $v$ over values, i.e., variables, $\unit$,
or variant values. 
A process can be
	an output $x\out {\tl v}.P$ which sends $\tilde v$ on $x$ and proceeds as $P$; 
	an input $x\inp {\tl y}.P$ which receives a sequence of values
	on $x$, substitutes them for $\tilde y$ in $P$;
	a parallel composition $P\pp Q$ of $P,Q$;
	replicated $*P$;
	a restriction $\res x P$ which creates a new channel $x$ and binds it in $P$;
	a $\picase{v}{x_i}{P_i}$ which offers a set of labelled processes, with labels being all different; or
	inaction $\nil$.
%
%

\smallpar{Standard \p types.}\label{sec:pi_types}
The syntax of \p types~\cite{Sangio01,KPT96} is defined in Figure~\ref{fig:pi_types}.
\begin{figure}[h!]
  \begin{displaymath}
    \begin{array}{l}
      \begin{array}{rcllllllll}
        \tau & ::=
        &\emptyset [] & \mbox{({no capability})}
        & \ |  & \ch [\wt T]& \mbox{(connection)} \\  
    	&& \ltp{i}{\wt T} & \mbox{(linear input)}
        &\ | &\ltp{o}{\wt T} & \mbox{(linear output)}
        & \ | & \ltp{\ch}{\wt T}  & \mbox{(linear connection)}\\[1mm]
        T  & ::= 
        & \tau & \mbox{({channel type})}
        &\ | & \variant {l_i}{T_i} & \mbox{(variant type)}
        &\ | & X & \mbox{(type var.)}\\
        & &\overline X& \mbox{(dual type var.)}
        &\ |& \unitT & \mbox{(unit type)}
        & \ |  &\rec X.T& \mbox{({recursive type})}
      \end{array}
    \end{array}
  \end{displaymath}
    \vspace{-1.5em}
  \caption{Standard $\pi$-calculus, types.}
  \label{fig:pi_types}
\end{figure}

\noindent
	$\tau$ ranges over channel types and
	$T$ over types. 
A channel type is
	a type with no capability $\emptyset []$, meaning it cannot be used further;
	a connection $\ch [\wt T]$, indefinitely used;
	a linear input $\ltp{i}{\wt T}$, a linear output $\ltp{o}{\wt T}$ or the combination of both,
	i.e., a linear connection $\ltp{\ch}{\wt T}$
	used  \emph{exactly once}~\cite{KPT96} according to its capability.
A type can be
	a channel type $\tau$;
	a variant $\variant {l_i}{T_i}$ being a set of labelled types, with labels being all different;
	a (dualised) type variable $X, \overline X$ or recursive type $\rec X.T$ or
	$\unitT$.
Again, we require recursive types to be {\em guarded} and use $\pt$ to denote the set of
	{\em closed} (no free type variables) and
	{\em guarded} standard \p types. 
To conclude, the definition of {\em unfolding} is the same as in the previous section.

\smallpar{On duality for linear types.}
Inspired by duality on session types, below we give the definition of
complement function $\picmplt{}$  and
co-inductive duality relation $\Pdual$ for linear \p types.
\begin{definition}[Complement function for linear \p types]
\label{fig:pi_dual}
\label{fig:pi_complement}
The complement function is defined as:
\[
\begin{array}{lcl@{\hskip 1em}lcl}
\picmplt{\ltp{i}{\wt T}} & = &\ltp{o}{\wt T}
&
\picmplt{X} & = & \overline X \\

\picmplt{\ltp{o}{\wt T}} & = & \ltp{i}{\wt T}
&
\picmplt{\overline X} & = & X \\

\picmplt{\rec X.T} & = & \rec X.{\picmplt{T\substIO{\rec{X}.{T}}{X}}}
&
\picmplt{\emp []} & = & \emp [] 
\end{array}
\]
\end{definition}
The definition of type substitution for linear types is the same as in the previous section,
where $\cmplt{}$ is replaced by $\picmplt$.
Before defining $\Pdual$, we give a co-inductive definition of subtyping and type equivalence for linear \p types.
For simplicity, we omit the subtyping on base types, standard channel types, or variant types
which can be found in the literature~\cite{Sangio01}.
\begin{definition}[Subtyping and type equivalence for linear \p types]
\label{def:pisubtyping}
A relation $\R \subseteq \pt \times \pt$ is a {\em type simulation} if $(T,S) \in \R$ implies the following:\\
$i)$ if $\unfold{T}=\emp []$ then $\unfold{S}=\emp []$\\
$ii)$ if $\unfold{T}=\ltp{i}{\wt T}$ then
			$\unfold{S}=\ltp{i}{\wt S} $ and $\wt{T} \R \wt{S}$\\
$iii)$ if $\unfold{T}=\ltp{o}{\wt T}$ then
			$\unfold{S}=\ltp{o}{\wt S} $ and $\wt{T} \R \wt{S}$\\
%
The {\em subtyping relation} $\sbt$ is defined by $T\pbt S$ if and only if there exists a type simulation $\R$
such that $(T,S) \in \R$.
The {\em type equivalence relation} $\pbteq$ is defined by $T\pbteq S$ if and only if
$T\pbt S$ and $S \pbt T$.
\end{definition}
\begin{definition}[Co-inductive duality for linear \p types]
\label{def:pi_dual}
A relation $\R \in \pt \times \pt$
is a {\em duality relation} if $(T,S)\in \R$ implies the following conditions:\\
$i)$ 	If $\unfold{T} = \emp []$ then $\unfold{S} = \emp []$\\
$ii)$ 	If $\unfold{T} = \ltp{i}{\wt T}$ then 
		$\unfold{S} = \ltp{o}{\wt S}$ and $\wt {T}\pbteq \wt S$\\
$iii)$ If $\unfold{T} =  \ltp{o}{\wt T}$ then 
		$\unfold{S} =  \ltp{i}{\wt T}$and $\wt {T}\pbteq \wt S$\\
The {\em co-inductive duality relation} $\Pdual$ is defined by $T\Pdual S$ iff
$\ \exists\ \R$, a duality relation such that  $(T,S) \in \R$.
\end{definition}

\begin{proposition}\label{prop:picmplt_dual}
Let $T, S\in \pt$, $\picmplt{T} = S\ \Longrightarrow\ T \Pdual S$.
\end{proposition}

%
\begin{proposition}[Idempotence]\label{prop:piidempotence}
Let $T,S,U\in \pt$. If $T\Pdual S$ and $S\Pdual U$ then $T \pbteq U$.
\end{proposition}
%
%
\begin{proposition}\label{prop:cmplt_idem}
Let $T,S,U\in \pt$. If $\picmplt T = S$ and $\picmplt S = U$ then $T \pbteq U$.
\end{proposition}

\section{Encoding recursive session types}
\label{sec:encoding}
Below we give the encoding of recursive session types into
recursive linear \p types and of session \p processes into standard \p processes.
It is based on the notions of: {\em linearity}, {\em variant types} and {\em continuation-passing} principle.
To preserve communication safety and privacy of session types,
we use linear channels.
To encode internal and external choice,
we adopt variant types and the \justcase process.
To preserve the sequentiality of session types and hence session fidelity,
we adopt the continuation-passing principle.

\smallpar{Types encoding.}
The encoding of session types is presented in Figure~\ref{fig:enc_types}.
Type $\nilT$ is encoded as the channel type with no capability $\emp [ ]$;
output ${\oc T.S}$ and input ${\wn T.S}$ session types are encoded as
linear output $\ltp{o}{\encT{T},\encT{\cmplt S}}$ and linear input  $\ltp{i}{\encT{T},\encT{S}}$ channel types
carrying the encoding of type $T$ and of continuation type $\cmplt S$ and $S$, respectively.
$\cmplt S$  is adopted in the output  since it is the type of a channel as seen by the receiver, namely the communicating counterpart.
Select ${ \select lS}$ and branch ${\branch lS}$
are encoded as linear output $\ltp{o}{\variant {l_i}{\encT{\cmplt{S_i}}}}$ and linear input $\ltp{i}{\variant {l_i}{\encT{S_i}}}$ types
carrying a variant type with the encoded continuation types; the reason for $\cmplt{S_i}$ is the same as before.
The encoding of a (dualised) type variable and a recursive session type is an homomorphism.

\smallpar{Terms encoding.}
The encoding of session \p terms, presented in Figure~\ref{fig:enc_terms}, uses a partial function
$f$ from variables to variables which performs a renaming of
linear variables into new linear variables
to respect their nature of being used exactly once and it is the identity function over standard variables.
The formal definition can be found in~\cite{D14Ext}.
\begin{figure*}[t!]
\centering
\begin{displaymath}
\begin{array}{llcl}
	\encT{\nilT}			&\defeq\  \emp [ ]							\\	
	\encT{\oc T.S}			&\defeq\  \ltp{o}{\encT{T},\encT{\cmplt S}}		\\	
	\encT{\wn T.S}			&\defeq\  \ltp{i}{\encT{T},\encT{S}}				\\	
	\encT{ \select lS}		&\defeq\  \ltp{o}{\variant {l_i}{\encT{\cmplt{S_i}}}}	\\	
	\encT{\branch lS}		&\defeq\  \ltp{i}{\variant {l_i}{\encT{S_i}}}			\\	
	\encT{X}				&\defeq\ X								\\	
	\encT{\overline X}		&\defeq\ \overline X							\\	
	\encT{\rec X.S}			&\defeq\ \rec X.{\encT{S}}							
\end{array}
\qquad\quad
\begin{array}{llcl}
    \encf{x}							&\defeq\ \ \f x				\\
    \encf{x\out v.P}					&\defeq\ \ \res c \f x\out {v,c}.\encx Pc			\\
    \encf{x\inp y.P}					&\defeq\ \ \f x\inp{y,c}.\encx Pc					\\
    \encf{\selection x {l_j}.P}			&\defeq\ \  \res c \f x\out {\vv{j}c}.\encx Pc			\\
    \encf{\branching xlP}				&\defeq\ \ \f x \inp y.\ \picase {y}{c}{\encx {P_i}c}	\\
    \encf{\res{xy}P}					&\defeq\ \ \res {c} \enc Pc						\\
    \encf{P\pp Q}					&\defeq\ \ \encf{P} \pp \encf{Q}					\\
    \encf{\res x P}					&\defeq\ \ \res x \encf P 						\\
    \encf{*P}						&\defeq\ \ *\encf{P}							\\
    \encf{\nil}						&\defeq\ \ \nil								\\
  \end{array}
\end{displaymath}
  \vspace{-1.5em}
\caption{Encoding of session types and terms}
\label{fig:enc_terms}
\label{fig:enc_types}
\end{figure*}
A variable $x$ is encoded as $\f x$;
an output ${x\out v.P}$ is encoded as an output on $\f x$ of $v$ and the
freshly created channel $c$ which replaces $x$ in the encoding of $P$.
The encoding of an input ${x\inp y.P}$ is an input on $\f x$  with placeholders $y$ and the continuation channel $c$ used in $\encx Pc$.
The encodings of selection ${\selection x {l_j}.P}$ and branching ${\branching xlP}$ are the output and input processes on
$\f x$, respectively.
The output carries a variant value ${\vv{j}c}$
where $l_j$ is the selected label and
$c$ the new channel to be used in the continuation.
The input has a continuation of a {\bf case}, offering the encoded processes of the branching.
The session restriction process ${\res{xy}P}$ is encoded as the restriction on $c$ which replaces
both of the endpoint $x,y$ in the encoding of $P$.
The rest of the equations states that the encoding of
parallel composition, standard channel restriction and replication is an homomorphism and
the encoding of the inaction process is the identity function.

\smallpar{Results of the encoding.}
The proofs of the following results can be found in~\cite{D14Ext}.
The following two lemmas relate
the encoding of equal and dual session types to equal and dual linear \p types.
%
%
\begin{lemma}[Encoding  $\sbteq$]\label{lem:equal_encoding}
$T, S\in \st$ and $T \sbteq S$.
If $\encT T = \tau$, then $\encT S= \sigma$ and $\tau \pbteq \sigma$.
\end{lemma}
\begin{lemma}[Encoding $\Sdual$]\label{lem:dual_encoding}
$T, S \in \st$ and $T \Sdual S$.
If $\encT{\unfold T}=\tau$ then, $\encT{{\unfold S}}= \sigma$ and $\tau \Pdual \sigma$. 
\end{lemma}
%
\begin{lemma} [Value Typing]\label{thm:sound_val}
	$\Gamma \vdash v:T$ if and only if $\encf{\Gamma} \vdash \encf{v}:\encT{T}$.
	\footnote{The encoding is extended to typing environments $\Gamma$ and the details can be checked in~\cite{D14Ext}.}
\end{lemma}
\begin{theorem} [Process Typing]\label{thm:sound_proc}
	$\Gamma \vdash P$ if and only if $\encf{\Gamma} \vdash \encf{P}$.
\end{theorem}

\begin{theorem}[Operational Correspondence]\label{thm:oc}
  Let $P$ be a session process.
  The following hold.
  \begin{enumerate}
  \item
	 If $P\to P'$ then $\encf P\to\hookrightarrow\encf {P'}$,
  \item 
	If $\encf P\to Q$ then,  $\exists\ P', \ctx E{}\cdot$ 
	such that $\ctx E{}P\to \ctx E{}{P'}$ and $Q\hookrightarrow\encT {P'}_{f'}$,
	where $f'$ is the updated $f$ after reduction
	and $\f x=\f y\ $ for all $\res{xy}\in\ctx E{}\cdot$.
  \end{enumerate}
\end{theorem}
$\hookrightarrow$ denotes $\equiv$
{possibly} extended with a {\bf case} reduction;
$\ctx E{}\cdot$ is an {\em evaluation context}.
\section{Example of encoding}
\label{sec:enc_example}
We present an error-free process which requires recursive session types.
We let $a,b$ range over standard channels and
$x,y,z,v, w$ range over session channels;
we associate a type to a variable in an object position in input or restriction as in~\cite{GH05}.
The typing rules and the operational semantics can be found in~\cite{D14Ext}.

Let 
$P= *\big(a\inp {x:T}. \selection xl.{a\out x}.\nil\big)$
be a replicated process that 
on a  standard channel $a$
receives a session channel $x$ on which
selects $l$ and proceeds as ${a\out x}$.
We have the following typing derivation for $P$:
$$
\infer[\text{T-Rep}]
{a : \ch T  \vdash *\big(a\inp {x:T}. \selection xl.{a\out x.\nil}\big)}
{
\infer[\text{T-In}]
{ a : \ch T  \vdash a\inp {x:T}. \selection xl.{a\out x}.\nil \qquad \un(a: \ch T)}
{
\infer[\text{T-Select}]
{ a : \ch T , x :  \oplus\{{l}: S \} \vdash \selection xl.{a\out x}.\nil \qquad  T \sbt \oplus \{{l}: S\}}
{
\infer[\text{T-Out}]
{ a : \ch T , x : S  \vdash {a\out x}.\nil}
{
\infer[\text{T-Nil}]
{a : \ch T \vdash \nil \qquad S \sbt T}
{}
}
}
}
}
$$
For this derivation to hold, $T$ and $S$ need to be such that
$T \sbt \oplus\{l:S\}$ and $S\sbt T$.
The simplest way to solve this system of subtyping in-equations is to have $S = T$,
which requires $T = \rec X. \oplus \{l:X\}$.

Let
$Q= *\big( b\inp{x:U}. x \triangleright\{l: b\out x.\nil\}\big)$;
it has dual behaviour to $P$
and the typing derivation is similar to above, with \textsc{T-Select} replaced by \textsc{T-Branch}.
We now have the in-equations $U\sbt \&\{l:S'\}$ and $S'\sbt U$.
We let
$U= \rec X.\&\{l:X\}$.
By Definition~\ref{def:Sdual}, we have $U\Sdual T$.
We now close $P$ and $Q$ with two auxiliary output processes, $a\out v.\nil$ and $b\out w.\nil$,
where $v,w$ are to be co-variables. Then, we have:
\begin{align*}
\emptyset \vdash
Sys
=&			\res{a:\ch T}\res{b:\ch U}\res {vw:T}\big(a\out v.\nil \pp b\out w.\nil \pp P \pp Q \big) \\
\equiv\to& 	\res{a:\ch T}\res{b:\ch U}\res {vw:T}\big(\selection vl.{a\out v}.\nil\pp w \triangleright\{l: b\out w.\nil\}\pp  P \pp Q\big)\\
\to& 			\res{a:\ch T}\res{b:\ch U}\res {vw:T}\big( a\out v.\nil \pp b\out w.\nil \pp P \pp Q \big)= Sys \mto
\end{align*}
The encoding of types is as follows.
Since
$U\Sdual T$ by Lemma~\ref{lem:dual_encoding} we have $\upsilon\Pdual \tau$.
\begin{align*}
\encT{U} &=\encT{\rec X.\&\{l:X\}}
= \rec X.\encT{\&\{l:X\}}
= \rec X. \ltp{i}{\langle {l}:{\encT{X}}\rangle}
=  \rec X. \ltp{i}{\langle {l}: {X}\rangle}= \upsilon\\
\encT{T} &= \encT{ \rec X. \oplus \{l:X\}}
= \rec X.\encT{  \oplus \{l:X\}}
=\rec X. \ltp{o}{\langle {l}:{\encT{\cmplt{X}}}\rangle}
= \rec X. \ltp{o}{\langle {l}:\overline {X}\rangle} = \tau
\end{align*}
Duality of session types boils down to opposite capabilities in the outermost level ($\ell_{\m i}, \ell_{\m o}$) and
the {\em same} carried type, where in~\cite{DGS12} same means {\em syntactic identity} and in the present means {\em type equivalence}.
{Unfolding is performed in order to test linear type duality and  the type equivalence of the carried type.}
\begin{align*}
\encf P& 	= \encf {*\big(a\inp {x}. \selection xl.{a\out x}.\nil\big)} 
		= *\encf{\big(a\inp {x}. \selection xl.{a\out x}.\nil\big)}
		= *\big( a\inp {x}. \encf{\selection xl.{a\out x}.\nil} \big)\\
	&	= *\big( a\inp {x}.  \res {c} x\out {\vv{}c}.\encx {{a\out x}.\nil}c \big)
		= *\big( a\inp {x}.  \res {c} x\out {\vv{}c}. {{a\out c}.\nil} \big)\\
\encf Q &	= \encf{*\big( b\inp{x}. x \triangleright\{l: b\out x.\nil\}\big)}
		= *\encf{\big( b\inp{x}. x \triangleright\{l: b\out x.\nil\}\big)}
		= *\big( b\inp{x}. \encf{x \triangleright\{l: b\out x.\nil\}\big)}\\
	&	= *\big( b\inp{x}. x\inp y.{\mathbf{case} \, {y} \, \mathbf{of}\, \{l\_c \triangleright \encx{b\out x.\nil\}}c \}}\big)
		= *\big( b\inp{x}. x\inp y.{\mathbf{case} \, {y} \, \mathbf{of}\, \{l\_c \triangleright {b\out c.\nil\}} \}}\big)
\end{align*}
\begin{align*}
\emptyset\vdash \encf{Sys}
		=&	\res{a}\res{b}\res {z}\encT {\big(a\out v.\nil \pp b\out w.\nil \pp P \pp Q \big)}_{f,\{v,w\mapsto z\}}
		=	\res{a}\res{b}\res {z}\big(a\out {z^+}.\nil \pp b\out{z^-}.\nil \pp \encT{P}_{f} \pp \encT{Q}_{f}  \big) \\
		\equiv&	\res{a}\res{b}\res {z}\big(a\out {z^+}.\nil \pp b\out {z^-}.\nil \pp
							a\inp {x}.  \res {c} x\out {\vv{}c}. {{a\out c}.\nil} \pp
							b\inp{x}. x\inp y.{\mathbf{case} \, {y} \, \mathbf{of}\, \{l\_c \triangleright {b\out c.\nil\}} \}}\pp \\
		&					 *\big( a\inp {x}.  \res {c} x\out {\vv{}c}. {{a\out c}.\nil} \big) \pp
							 *\big( b\inp{x}. x\inp y.{\mathbf{case} \, {y} \, \mathbf{of}\, \{l\_c \triangleright {b\out c.\nil\}} \}}\big) \big)
\end{align*}
\begin{align*}
		\to&	\res{a}\res{b}\res {z}\big(
							  \res {c}{z^+}\out {\vv{}c^-}. {{a\out {c^+}}.\nil} \pp
							  {z^-}\inp y.{\mathbf{case} \, {y} \, \mathbf{of}\, \{l\_c^- \triangleright {b\out {c^-}.\nil\}} \}} \pp \encf P\pp \encf Q\big)\\
		\to&	\res{a}\res{b}\res {z}\big(
							  \res {c}. {{a\out {c^+}}.\nil} \pp
							  {\mathbf{case} \, {\vv{}c^-} \, \mathbf{of}\, \{l\_c^- \triangleright {b\out {c^-}.\nil\}} \}} \pp \encf P\pp \encf Q\big) \\
		\hookrightarrow&	\res{a}\res{b}\big(
							  \res {c}. {{a\out {c^+}}.\nil} \pp
							  {{b\out {c^-}.\nil\}} \}} \pp \encf P\pp \encf Q\big)\equiv \encf{Sys}\mto\\ 
\end{align*}

\section{Conclusions and Future Work}
\label{sec:concl}
In this paper we present an encoding of recursive session types
into recursive linear types and session processes into corresponding \p processes.
The encoding is a conservative extension of the one given in~\cite{DGS12}.
It uses $\cmplt{}$ instead of $\ \overline\cdot\ $, because the latter is inadequate in the
presence of recursive types~\cite{BH13,BDGK14}.
Since these two functions coincide for finite session types,
the encoding in~\cite{DGS12} remains sound.
We prove the faithfulness of the present encoding with respect to typing derivations and operational semantics,
following the same line of~\cite{DGS12,DardhaPhDThesis}.
As long as future work is concerned,
we would like to test our encoding under different dualities for sessions presented in~\cite{BDGK14}.
Moreover, as in~\cite{DGS12} we would like to extend
the present encoding to advanced features like polymorphism or higher-order, or multiparty session types.

\noindent
{\bf Acknowledgements.}
The author would like to thank Simon J. Gay, Elena Giachino and Davide Sangiorgi for the inspiring and very useful discussions.

\bibliography{my_biblio}
\bibliographystyle{eptcs}

\end{document}